\documentclass[fleqn,10pt]{wlscirep}

\newcommand{\msol}{\ensuremath{\, {\rm M}_\odot}}

\newcommand{\mgas}{{M}_{\rm gas}}   
\newcommand{\mstar}{{M}_{\rm star}}
\newcommand{\mhalo}{{M}_{\rm halo}}

\newcommand*{\Mwl}{$M_{\rm WL}$}
\newcommand*{\Lxrass}{$L_{X,\rm RASS}$}
\newcommand*{\Lx}{$L_{X,\rm ce}$}
\newcommand*{\Tx}{$T_{X,\rm ce}$}
\newcommand*{\Mgas}{$M_{\rm gas}$}
\newcommand*{\Mstar}{$M_{\rm star}$}
\newcommand*{\Yx}{$Y_{X}$}
\newcommand*{\Ysza}{$Y_{\rm SZA}$}
\newcommand*{\Ypl}{$Y_{\rm Pl}$}
\newcommand*{\Lk}{$L_{K,\rm tot}$}
\newcommand*{\Lbcg}{$L_{K,\rm BCG}$}
\newcommand*{\richness}{$\lambda$}

\usepackage[normalem]{ulem}
\usepackage{fancyhdr}
\usepackage{tabu}
\usepackage{xcolor,colortbl}
\usepackage{subcaption}

\newcolumntype{a}{>{\columncolor{Gray}}l}
\newcolumntype{r}{>{\columncolor[HTML]{FFCCC9}}c}
\newcolumntype{b}{>{\columncolor[HTML]{F0E68C}}c}

\title{Detection of anti-correlation of hot and cold baryons in galaxy clusters}

\author[1,2,*]{Arya Farahi}
\author[3]{Sarah L. Mulroy}
\author[1,4]{August E. Evrard}
\author[3]{Graham P. Smith}
\author[5,6]{Alexis Finoguenov}
\author[7,8]{Herv\'e Bourdin}
\author[9]{John E. Carlstrom}
\author[10]{Chris P. Haines}
\author[11]{Daniel P. Marrone}
\author[8]{Rossella Martino}
\author[7,8]{Pasquale Mazzotta}
\author[11]{Christine O'Donnell}
\author[12,13,14]{Nobuhiro Okabe}
\affil[1]{Department of Physics, University of Michigan, Ann Arbor, MI 48109, USA}
\affil[2]{McWilliams Center for Cosmology, Department of Physics, Carnegie Mellon University, Pittsburgh, PA 15213, USA}
\affil[3]{School of Physics and Astronomy, University of Birmingham, Birmingham, B15 2TT, England}
\affil[4]{Department of Astronomy, University of Michigan, Ann Arbor, MI 48109, USA}
\affil[5]{Department of Physics, University of Helsinki, Gustaf H\"allstr\"omin katu 2a, 00014 Helsinki, Finland}
\affil[6]{Max-Planck-Institute for Extraterrestrial Physics, Giessenbachstrasse, 85741 Garching, Germany}
\affil[7]{Harvard Smithsonian Centre for Astrophysics, 60 Garden Street, Cambridge, MA 02138, USA}
\affil[8]{Dipartimento di Fisica, Universit\`a degli Studi di Roma ``Tor Vergata'', via della Ricerca Scientifica 1, 00133, Roma, Italy}
\affil[9]{Kavli Institute for Cosmological Physics, Department of Astronomy and Astrophysics, University of Chicago, IL 60637}
\affil[10]{INAF - Osservatorio Astronomico di Brera, Via Brera 28, 20122 Milano, Italy}
\affil[11]{Steward Observatory, University of Arizona, 933 North Cherry Avenue, Tucson, AZ 85721, USA}
\affil[12]{Department of Physical Science, Hiroshima University, 1-3-1 Kagamiyama, Higashi-Hiroshima, Hiroshima 739-8526, Japan}
\affil[13]{Hiroshima Astrophysical Science Center, Hiroshima University, 1-3-1 Kagamiyama, Higashi-Hiroshima, Hiroshima 739-8526, Japan}
\affil[14]{Core Research for Energetic Universe, Hiroshima University, 1-3-1 Kagamiyama, Higashi-Hiroshima, Hiroshima 739-8526, Japan}
\affil[*]{aryaf@umich.edu}

\begin{abstract}
The largest clusters of galaxies in the Universe contain vast amounts of dark matter, plus baryonic matter in two principal phases, a majority hot gas component and a minority cold stellar phase comprising stars, compact objects, and low-temperature gas. Hydrodynamic simulations indicate that the highest-mass systems retain the cosmic fraction of baryons, a natural consequence of which is anti-correlation between the masses of hot gas and stars within dark matter halos of fixed total mass.  We report observational detection of this anti-correlation based on 4 elements of a $9\times9$-element covariance matrix for nine cluster properties, measured from X-ray, optical, infrared and millimetre wavelength observations of 41 clusters from the Local Cluster Substructure Survey. These clusters were selected using explicit and quantitative selection rules that were then encoded in our hierarchical bayesian model. Our detection of anti-correlation is consistent with predictions from contemporary hydrodynamic cosmological simulations that were not tuned to reproduce this signal.
\end{abstract}
\begin{document}

\flushbottom
\maketitle

\thispagestyle{empty}

\section*{Introduction}

Dark matter, whose nature remains elusive, and ordinary matter described by the Standard Model of particle physics, are the strongly clustered materials of our Universe, with the latter component referred to as baryonic matter, or more simply baryons, by observational cosmologists\cite{Dodelson:cosmo2003}. The question of how well these two components trace one another, across spatial scales and cosmic time, is central to our understanding of the astrophysics that drives galaxy formation and offers clues to the thermal nature of dark matter and other new physics.

Assuming weak-field (Newtonian) gravitational accretion and collisional shocks under the approximation of spherical symmetry, self-similar solutions \cite{Bertschinger:1985} emerge in which both collisionless dark matter and collisional baryonic fluids develop similar radial profiles when expressed in terms of a characteristic physical radius, the turn-around radius \cite{GunnGott:1972} at which the perturbed Hubble flow is stationary. A key implication of this simple model is that dark matter and baryons exhibit no radial separation.  Collapsed structures, referred to as halos, should retain the cosmic mix of these different fluids at all radii.

The most massive dark matter halos host groups and clusters of galaxies. Early X-ray measurements of the hot gas content of clusters upended the standard cold dark matter (CDM) model orthodoxy of a matter dominated universe\cite{White:1993} before observations of Type Ia supernovae ushered in the current $\Lambda$CDM model of a universe dominated by a smooth dark energy component\cite{Riess:1998}. The argument against a matter dominated universe relied on a fair sampling hypothesis, namely that the mean baryon fraction within clusters (ratio of baryonic mass to total mass) accurately reflects the cosmic mean baryon fraction.  A natural consequence of this hypothesis is that the hot and cold fractions of baryons in clusters should be anti-correlated; at fixed total mass, clusters with more cold baryons should have less hot baryons, and vice versa. While this model does not define how baryons are partitioned into these phases, the constancy of the sum implies that a particular system with more hot gas than average must contain a lower stellar mass than average, and vice-versa.  

However, this simple model ignores important non-spherical and non-gravitational effects such as hierarchical mergers driven by large-scale filaments and the redistribution of energy, momentum and mass (generically termed feedback) by supernovae and active galactic nuclei (AGN). In low mass halos that host only one bright galaxy like the Milky Way, feedback is energetic enough to vent hot gas phase baryons out of these relatively shallow gravitational potentials\cite{SilkMamon:2012}.  Even smaller halos of dwarf or satellite galaxies suffer severe baryon losses from collective supernova explosions.  At the other extreme, the massive halos of rich galaxy clusters have much deeper gravitational potentials that shield them from feedback-driven baryon venting outside of their core regions\cite{Voit:2017}. Thus, clusters are likely to be closed baryon boxes, unbiased reservoirs of the cosmic baryon fraction. 

Studies of mean trends in gas and stellar mass fraction\cite{Vikhlinin:2006,Gonzalez:2013} support the expectation that massive clusters are more closed than smaller halos, based on the trend of increasing baryon fraction with halo mass. However, measurements of absolute baryon fractions are currently subject to uncertain biases of ${\cal O}(10\%)$ in estimates of total mass, and this systematic uncertainty limits the reliability of comparison with the cosmic mean baryon fraction. We take a complementary approach based on {\sl variance about mean behavior}, particularly the covariance of hot gas mass and stellar mass conditioned on total mass. This approach is encouraged by recent findings of strongly negative correlation coefficients ($r \lesssim - 0.5$) from a pair of complex, multi-fluid cosmological simulations in which this statistic has been measured\cite{Wu:2015,Farahi:2017bahamas}.

Observational studies have explored baryonic properties conditioned on estimated halo mass, particularly X-ray and thermal Sunyaev-Zel'dovich (SZ) Effect\cite{SZ:1980} signatures from the hot gas phase and optical/infrared properties of galaxies\cite{Giodini:2013}.  While correlations among internal hot gas properties have been measured in a few empirical studies \cite{Mantz:2010scaling, Maughan:2014, Mantz:2016Scaling, Andreon:2017}, the degree of correlation between hot gas and galactic components has not yet been investigated. The minimum requirement for such an analysis is to obtain high quality observations of both stellar and gas properties for a cluster sample with well-defined selection rules and robust estimates of total cluster mass. Currently, these requirements are only fulfilled by the Local Cluster Substructure Survey (LoCuSS), a multi-wavelength survey of the 41 X-ray brightest galaxy clusters at redshifts of $0.15 < z < 0.3$.

The LoCuSS sample is selected by applying a redshift-dependent X-ray luminosity ($L_X$) cut to clusters identified in the \emph{ROSAT} All-sky Survey (RASS) catalog at high galactic latitudes. The multi-wavelength observations used in this study, obtained over the period of a decade (2005-2014) by co-authors, includes optical imaging data from the Subaru 8.2-m telescope, infrared data from the 3.8-m United Kingdom Infrared Telescope on Mauna Kea (UKIRT), X-ray observations from the \emph{Chandra} and \emph{XMM-Newton} satellites, and millimeter observations of the thermal SZ effect from the Planck satellite and the Sunyaev-Zel'dovich Array (SZA).

Here we report the first observational detection of anti-correlation between the hot and cold baryon contents of galaxy clusters. The key measurements are a subset of posterior estimates of 36 pairwise correlations among nine cluster properties derived from the LoCuSS observations, most measured within a radial scale defined by the weak-lensing estimate of each system's mass\cite{Okabe:2016}. Details of the galaxy cluster sample and posterior measures of the slope, variance, and \Lxrass--property covariance for nine properties are presented in a companion work\cite{Mulroy:inprep}. Our detection of anti-correlation supports independent evidence that massive galaxy clusters retain close to the cosmic mix of baryons and dark matter, a finding that can underpin improved cluster cosmology from cross-wavelength sample analysis.

\section*{Results}

Table \ref{tab:obs} lists the observable properties employed in this analysis.  We model the data using a likelihood based on log-normal property covariance about mean scaling relations behaving as power-laws in halo mass\cite{Evrard:2014}.  We employ default redshift scaling behaviors\cite{Kaiser1986}, but this assumption is unimportant due to the narrow redshift range of the sample.  We assume that, on average, weak gravitational lensing measurements provide  unbiased estimates of true cluster masses with 0.2 fractional scatter. To model the selection effect, we employ the threshold selection condition for \Lxrass\ emission used to define the LoCuSS cluster sample. X-ray emission offers clearer identification of massive halos, being less prone than cluster properties measured at other wavelengths to confusion from additional halos projected along the line of sight.  While imprecise models of sample selection can bias scaling parameter estimates, we show in the Supplementary Material that inferred correlations among property pairs are insensitive to biases in posterior slope and variance estimates (See Supplementary Figures \ref{fig:corr-test-1} and \ref{fig:corr-test-2}). 

\begin{table*}
\caption{Non-selection elements of the property vector of the LoCuSS cluster sample. Each is an integated observed quantity or composite thereof.  The full analysis includes the original selection property, \Lxrass, described in the companion paper\cite{Mulroy:inprep}. The subscript 500 below is the multiple of the universe's critical density, $\rho_{\rm crit}(z)$) employed in the the enclosed density condition used to define physical size of cluster, $M_{500} \equiv M(< R_{500}) = (\frac{4\pi}{3}) 500 \rho_{\rm crit}(z)  R_{500}^3 $. Size estimates are derived from weak-lensing (WL) measurements. Ref. \cite{Mulroy:inprep} provides details of the instruments and methods used for each property. } \label {tab:obs}
	\begin{center}
		\tabcolsep=0.8mm
		\begin{tabular}{ | l | l | l | }
            \hline
            Element & Unit & Description \\
            \hline
\Lx$E^{-1}(z)$ & $ 10^{44} \, \rm erg\,s^{-1} $ & core-excised, bolometric X-ray luminosity \\
$ \rm k_B$ \Tx  & keV & core-excised intracluster medium (ICM) thermal temperature \\ 
\Mgas$E(z)$ & $10^{14} \,\, M_{\odot}$ & ICM gas mass within WL $R_{500}$\\ 
\Yx$E(z)$ & $10^{14} \,\, M_{\odot}\rm \,\, keV$ & ICM (spherical) X-ray thermal energy within WL $R_{500}$\\ 
\Ysza$E(z)$ & $10^{-5} \,\, \rm Mpc^2$ & ICM (spherical) SZ amplitude  \\ 
\Ypl$E(z)$ & $10^{-5} \,\, \rm Mpc^2$ & ICM (cylindrical) SZ amplitude \\ 
\Lbcg$E(z)$ & $10^{12} \,\, L_{\odot}$ & Brightest Cluster Galaxy (BCG) $K$-band luminosity \\ 
\Lk$E(z)$ & $10^{12} \,\, L_{\odot}$ & total $K$-band luminosity within WL $R_{500}$\\ 
\richness$E(z)$ & none & redMaPPer richness (count of galaxies) \\
$M_{\rm WL}$ $E(z)$ & $10^{14} \,\, M_{\odot}$ & weak-lensing estimate of mass, $M_{500}$ \\
            \hline
        \end{tabular}
	\end{center}
{\footnotesize}
\end{table*}

\begin{table*}[h!]
\centering
\caption{ {\bf Lower Triangle}:  Posterior median and 68$^{\rm th}$ percentile range of property pair correlation coefficients at fixed weak-lensing mass. Color encodes the magnitude and sign of the correlation coefficient, with red (blue) showing positive (negative) values. {\bf Upper Triangle}: Statistical significance (p-value) of the sign of the estimated property correlation, calculated as the cumulative posterior probability of having positive (negative) correlation values if the median is negative (positive). {\bf Diagonal}: Posterior median and 68$^{\rm th}$ percentile range of the intrinsic scatter of each property at fixed weak-lensing mass.   }
\label{tab:corr-post}{\footnotesize \tabulinesep=1.2mm
\begin{tabu}{|>{\columncolor[HTML]{A9A9A9}}l|c|c|c|c|c|c|c|c|c|}
\hline
\rowcolor[HTML]{A9A9A9} \cellcolor[HTML]{333333}  & \Lx\ & \Mgas\ & \Tx\ & \Yx\ & \Ypl\ & \Ysza\ & \Lbcg\ & \richness\ & \Lk\  \\ \hline
\Lx\  & \cellcolor[HTML]{CCCCCD}  $0.36 ^{+0.05 }_{-0.04 }$  & \cellcolor[rgb]{0.85, 0.95, 0.85}  $0.001$  & \cellcolor[rgb]{0.85, 0.95, 0.85}  $0.001$  & \cellcolor[rgb]{0.85, 0.95, 0.85}  $<10^{-3}$  & \cellcolor[rgb]{0.85, 0.95, 0.85}  $0.007$  & \cellcolor[rgb]{0.85, 0.95, 0.85}  $0.03$  & \cellcolor[rgb]{0.85, 0.95, 0.85}  $0.09$  & \cellcolor[rgb]{0.85, 0.95, 0.85}  $0.09$  & \cellcolor[rgb]{0.85, 0.95, 0.85}  $0.018$  \\ \hline
\Mgas\  & \cellcolor[rgb]{1.000, 0.469, 0.241}  $ 0.76 ^{+0.09 }_{-0.13 }$  & \cellcolor[HTML]{CCCCCD}  $0.15 ^{+0.03 }_{-0.03 }$  & \cellcolor[rgb]{0.85, 0.95, 0.85}  $0.28$  & \cellcolor[rgb]{0.85, 0.95, 0.85}  $0.008$  & \cellcolor[rgb]{0.85, 0.95, 0.85}  $0.011$  & \cellcolor[rgb]{0.85, 0.95, 0.85}  $0.37$  & \cellcolor[rgb]{0.85, 0.95, 0.85}  $0.46$  & \cellcolor[rgb]{0.85, 0.95, 0.85}  $0.30$  & \cellcolor[rgb]{0.85, 0.95, 0.85}  $0.06$  \\ \hline
\Tx\  & \cellcolor[rgb]{1.000, 0.657, 0.510}  $ 0.49 ^{+0.13 }_{-0.16 }$  & \cellcolor[rgb]{1.000, 0.910, 0.871}  $ 0.13 ^{+0.20 }_{-0.22 }$  & \cellcolor[HTML]{CCCCCD}  $0.19 ^{+0.04 }_{-0.03 }$  & \cellcolor[rgb]{0.85, 0.95, 0.85}  $<10^{-3}$  & \cellcolor[rgb]{0.85, 0.95, 0.85}  $0.22$  & \cellcolor[rgb]{0.85, 0.95, 0.85}  $0.06$  & \cellcolor[rgb]{0.85, 0.95, 0.85}  $0.36$  & \cellcolor[rgb]{0.85, 0.95, 0.85}  $0.16$  & \cellcolor[rgb]{0.85, 0.95, 0.85}  $0.07$  \\ \hline
\Yx\  & \cellcolor[rgb]{1.000, 0.404, 0.148}  $ 0.85 ^{+0.07 }_{-0.09 }$  & \cellcolor[rgb]{1.000, 0.584, 0.406}  $ 0.59 ^{+0.14 }_{-0.19 }$  & \cellcolor[rgb]{1.000, 0.529, 0.327}  $ 0.67 ^{+0.11 }_{-0.14 }$  & \cellcolor[HTML]{CCCCCD}  $0.32 ^{+0.05 }_{-0.05 }$  & \cellcolor[rgb]{0.85, 0.95, 0.85}  $0.003$  & \cellcolor[rgb]{0.85, 0.95, 0.85}  $0.04$  & \cellcolor[rgb]{0.85, 0.95, 0.85}  $0.10$  & \cellcolor[rgb]{0.85, 0.95, 0.85}  $0.33$  & \cellcolor[rgb]{0.85, 0.95, 0.85}  $0.04$  \\ \hline
\Ypl\  & \cellcolor[rgb]{1.000, 0.622, 0.460}  $ 0.54 ^{+0.12 }_{-0.16 }$  & \cellcolor[rgb]{1.000, 0.605, 0.436}  $ 0.56 ^{+0.13 }_{-0.18 }$  & \cellcolor[rgb]{1.000, 0.893, 0.847}  $ 0.15 ^{+0.18 }_{-0.20 }$  & \cellcolor[rgb]{1.000, 0.580, 0.400}  $ 0.60 ^{+0.12 }_{-0.16 }$  & \cellcolor[HTML]{CCCCCD}  $0.28 ^{+0.04 }_{-0.04 }$  & \cellcolor[rgb]{0.85, 0.95, 0.85}  $0.47$  & \cellcolor[rgb]{0.85, 0.95, 0.85}  $0.21$  & \cellcolor[rgb]{0.85, 0.95, 0.85}  $0.14$  & \cellcolor[rgb]{0.85, 0.95, 0.85}  $0.43$  \\ \hline
\Ysza\  & \cellcolor[rgb]{1.000, 0.712, 0.588}  $ 0.41 ^{+0.16 }_{-0.20 }$  & \cellcolor[rgb]{1.000, 0.947, 0.924}  $ 0.08 ^{+0.21 }_{-0.25 }$  & \cellcolor[rgb]{1.000, 0.751, 0.644}  $ 0.36 ^{+0.20 }_{-0.22 }$  & \cellcolor[rgb]{1.000, 0.725, 0.607}  $ 0.39 ^{+0.17 }_{-0.21 }$  & \cellcolor[rgb]{0.984, 0.988, 1.000}  $ -0.02 ^{+0.21 }_{-0.22 }$  & \cellcolor[HTML]{CCCCCD}  $0.29 ^{+0.08 }_{-0.07 }$  & \cellcolor[rgb]{0.85, 0.95, 0.85}  $0.40$  & \cellcolor[rgb]{0.85, 0.95, 0.85}  $0.32$  & \cellcolor[rgb]{0.85, 0.95, 0.85}  $0.25$  \\ \hline
\Lbcg\  & \cellcolor[rgb]{1.000, 0.864, 0.806}  $ 0.19 ^{+0.13 }_{-0.14 }$  & \cellcolor[rgb]{0.980, 0.986, 1.000}  $ -0.02 ^{+0.18 }_{-0.19 }$  & \cellcolor[rgb]{1.000, 0.956, 0.937}  $ 0.06 ^{+0.17 }_{-0.17 }$  & \cellcolor[rgb]{1.000, 0.856, 0.794}  $ 0.21 ^{+0.15 }_{-0.16 }$  & \cellcolor[rgb]{1.000, 0.911, 0.873}  $ 0.13 ^{+0.16 }_{-0.16 }$  & \cellcolor[rgb]{1.000, 0.962, 0.946}  $ 0.05 ^{+0.20 }_{-0.21 }$  & \cellcolor[HTML]{CCCCCD}  $0.34 ^{+0.05 }_{-0.04 }$  & \cellcolor[rgb]{0.85, 0.95, 0.85}  $0.17$  & \cellcolor[rgb]{0.85, 0.95, 0.85}  $0.43$  \\ \hline
\richness\  & \cellcolor[rgb]{0.764, 0.835, 1.000}  $ -0.24 ^{+0.18 }_{-0.17 }$  & \cellcolor[rgb]{0.893, 0.925, 1.000}  $ -0.11 ^{+0.21 }_{-0.21 }$  & \cellcolor[rgb]{0.761, 0.833, 1.000}  $ -0.24 ^{+0.24 }_{-0.23 }$  & \cellcolor[rgb]{0.917, 0.942, 1.000}  $ -0.08 ^{+0.19 }_{-0.20 }$  & \cellcolor[rgb]{1.000, 0.859, 0.798}  $ 0.20 ^{+0.16 }_{-0.19 }$  & \cellcolor[rgb]{0.879, 0.915, 1.000}  $ -0.12 ^{+0.25 }_{-0.25 }$  & \cellcolor[rgb]{0.829, 0.880, 1.000}  $ -0.17 ^{+0.18 }_{-0.18 }$  & \cellcolor[HTML]{CCCCCD}  $0.25 ^{+0.05 }_{-0.04 }$  & \cellcolor[rgb]{0.85, 0.95, 0.85}  $0.014$  \\ \hline
\Lk\  & \cellcolor[rgb]{0.411, 0.588, 1.000}  $ -0.59 ^{+0.27 }_{-0.22 }$  & \cellcolor[rgb]{0.438, 0.607, 1.000}  $ -0.56 ^{+0.36 }_{-0.28 }$  & \cellcolor[rgb]{0.521, 0.665, 1.000}  $ -0.48 ^{+0.32 }_{-0.29 }$  & \cellcolor[rgb]{0.470, 0.629, 1.000}  $ -0.53 ^{+0.30 }_{-0.26 }$  & \cellcolor[rgb]{1.000, 0.964, 0.948}  $ 0.05 ^{+0.29 }_{-0.32 }$  & \cellcolor[rgb]{0.744, 0.821, 1.000}  $ -0.26 ^{+0.37 }_{-0.39 }$  & \cellcolor[rgb]{1.000, 0.961, 0.945}  $ 0.06 ^{+0.32 }_{-0.32 }$  & \cellcolor[rgb]{1.000, 0.461, 0.230}  $ 0.77 ^{+0.16 }_{-0.27 }$  & \cellcolor[HTML]{CCCCCD}  $0.09 ^{+0.05 }_{-0.03 }$  \\ \hline
\end{tabu}}
\end{table*}

\begin{figure}[h!]
  \centering
  \includegraphics[width=0.8\linewidth]{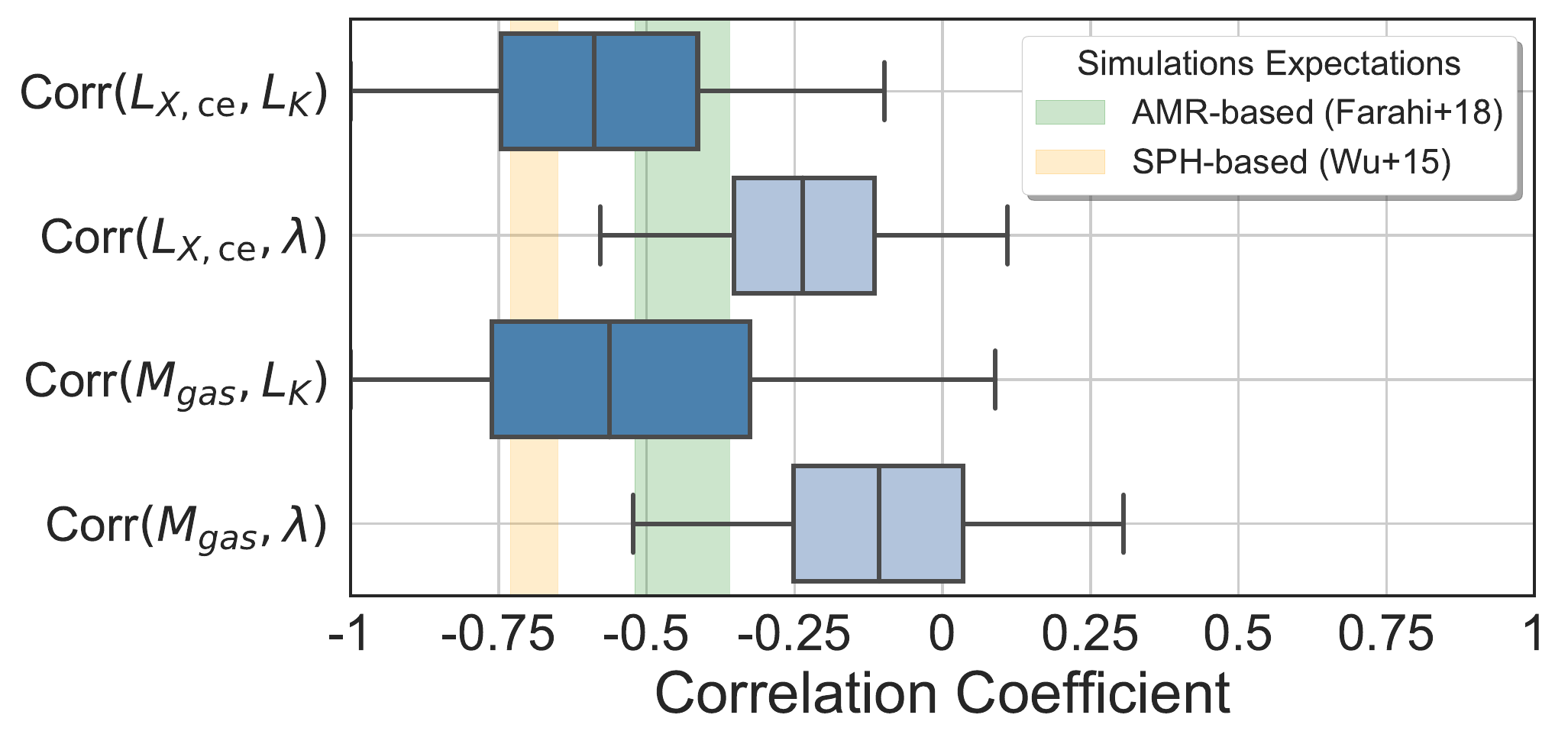} 
 \caption{Correlation coefficients of pairs of hot gas and galactic properties at fixed weak-lensing mass. Empirical results from the LoCuSS cluster sample are shown as box plots, and the background bands show stellar mass and hot gas mass correlation coefficients from  two independent hydrodynamical simulations, the Adaptive Mesh Refinement (AMR)-based Rhapsody-G \cite{Wu:2015} and the Smoothed Particle Hydrodynamics (SPH)-based BAHAMAS+MACSIS simulations \cite{Farahi:2017bahamas}.  In the box plot representation, the middle line shows the median of the posterior distribution while box edges show the first and third quartiles (25$^{\rm th}$ and 75$^{\rm th}$ percentiles, respectively) with whiskers extending to show the inner 95\% of the posterior distribution (2.5\% to 97.5\%). The shaded regions are $68\%$ confidence intervals.  }
 \label{fig:corr-x-ray-optical}
\end{figure}

Our analysis takes a hierarchical Bayesian approach that accounts for the effects of the sample selection, measurement error covariance induced by the use of a common sky aperture and other effects, as well as the halo space density as a function of mass for a $\Lambda$CDM cosmology.  Uninformative priors are used in the regression; all quoted constraints are derived solely from the sample data.  Our model simultaneously constrains the population scaling parameters associated with the multi-wavelength ensemble, the slopes, normalizations, and the mass-conditioned property covariance (see equation~(\ref{eq:cov-estimator}) in the Methods section).  We report pairwise correlation coefficients, i.e. the covariance divided by the intrinsic scatter of each observable, as in equation~(\ref{eq:r-estimator}).

An analysis of variance must be cognizant of astrophysical sources of scatter extrinsic to the host halos of the cluster sample.  In particular, other halos along the line-of-sight will add correlated noise to some of a cluster's observed properties\cite{NohCohn:2012}.  In the Supplementary Figure \ref{fig:projection-test}, we show that such sources of systematic error, including  projection, tend to dilute the magnitude of an intrinsically anti-correlated property pair.  We argue that it is conservative, then, to consider the measured magnitude of an anti-correlation between stellar mass and hot gas mass as effectively a lower limit to the underlying halo population value.

Table \ref{tab:corr-post} presents the full property covariance matrix at fixed weak-lensing mass derived from the LoCuSS sample.  While we report the entire matrix, our focus is mainly on the last two rows and columns that contain optical properties.  The lower triangle elements summarize the correlation coefficients of property pairs while the diagonal elements provide standard deviations of each property.  Median values from the Markov chain Monte Carlo (MCMC) chains, $\sim 10^5$ in length, are quoted, and uncertainties give 68\% confidence limits.  As explained below in Methods, we impose a minimum value of $0.05$ on the intrinsic scatter in the log of K-band luminosity, $\ln$ \Lk, at fixed halo mass when determining these statistics. The upper triangle gives the odds that each element has a sign opposite to that of its median value.

We first note the physically sensible result that the two independently measured properties reflecting a halo's stellar mass -- total $K$--band luminosity, \Lk, and optical richness \richness\ -- have a strong positive correlation, $r = 0.77^{+0.16}_{-0.27}$.  The probability of this value being negative is very small, $1.4\%$.  Of these two measures, the quantity \richness\ appears noisier, with median intrinsic scatter of $25\%$ compared to only $9\%$ for \Lk, but this may also reflect the different measurement errors quoted for each property.  The fractional statistical uncertainties in \richness\ are a factor $\sim 3$ smaller than those in \Lk. As we show in the Supplementary Figure \ref{fig:corr-test-2}, bias and/or extra noise relative to the underlying halo population statistics will tend to dilute measured (anti-)correlations, and these effects can explain why some galaxy-hot gas property pairs yield weaker evidence of anti-correlation.

In the bottom two rows of Table~\ref{tab:corr-post}, the elements linking galaxy and hot gas properties are mainly negative.  Figure~\ref{fig:corr-x-ray-optical} highlights the correlation coefficients between the galaxy measures and two key measures of hot gas: the core-excised X-ray luminosity and the derived gas mass.  Boxes show inner quartiles ($25-75$-percentile) and whiskers encompass the inner $95$ percent of the marginalized posterior distributions. All pairs tend to be negative, as anticipated by the correlations between hot gas mass and stellar mass seen in hydrodynamical simulations \cite{Wu:2015, Farahi:2017bahamas}, shown as background bands in Figure~\ref{fig:corr-x-ray-optical}. The consistency in sign of hot-cold phase covariance elements between observed proxy measures and their simulation-derived counterparts is an encouraging sign of fidelity in the sophisticated astrophysical treatments employed to model the coupled evolution of multiple baryon components at sub-cluster scales in modern cosmological simulations\cite{BorganiKravtsov:2011}.  A consistent feature of such simulations is that the mean baryon fraction measured within the characteristic $R_{500}$ length scale used in this work approaches the cosmic value as system mass increases.  This aspect, along with a reduction in the population variance in baryon component mass fractions, supports the fair sampling hypothesis and allows the most massive clusters to serve as cosmic distance rulers\cite{Allen:2004}.

Due to the modest sample size, the uncertainty on any individual correlation coefficient remains large. Examining the upper triangle of Table~\ref{tab:corr-post}, we find that the pairing of \Lk\ and \Lx\ is the strongest indicator of anti-correlation, with only a $1.8\%$ chance of being zero or positive.  As noted in Fig.~\ref{fig:p-val-test} of the Methods section, the odds of a positive \Lk--\Lx\ correlation drop below one percent if the intrinsic scatter in \Lk\ at fixed halo mass is larger than 0.07.  For the \richness\ measure of stellar mass, the evidence is somewhat weaker, with a 9\% chance that it correlates positively with \Lx\ at fixed halo mass. 

It has previously been argued\cite{Lin:2004,Mulroy2014} that the $K$-band integrated light is a more accurate indicator of a cluster's total stellar mass than the number of optically-selected galaxies, i.e.\ optical richness.  Our results appear to reinforce this finding, as the anti-correlations for \Lk\ and X-ray properties are systematically more negative than those inferred using \richness.  But, as noted above, underestimation of the statistical uncertainty in \richness\ could also play a role in diluting \richness\ correlations\cite{Costanzi:2018}  (see Supplementary Figures \ref{fig:corr-test-2} and \ref{fig:projection-test}).

The findings of anti-correlation using core-excised X-ray luminosity are reinforced by the derived gas mass, \Mgas. Again, the infrared light provides a tighter constraint, with only a $6\%$ chance of being zero or positive, while the odds rise to $30\%$ when using \richness.  In the companion paper, we note that the slope of the \Mgas\ scaling with halo mass is $\sim 2.5\sigma$ lower than values derived by previous studies based solely on X-ray observations and also slopes inferred from modern hydrodynamic simulations. A bias in slope could dilute the anti-correlation signal and explain why \Lx\ provides more significant evidence of anti-correlation (Supplementary Figure \ref{fig:corr-test-1}).

\begin{figure}
  \centering
  \includegraphics[width=0.8\linewidth]{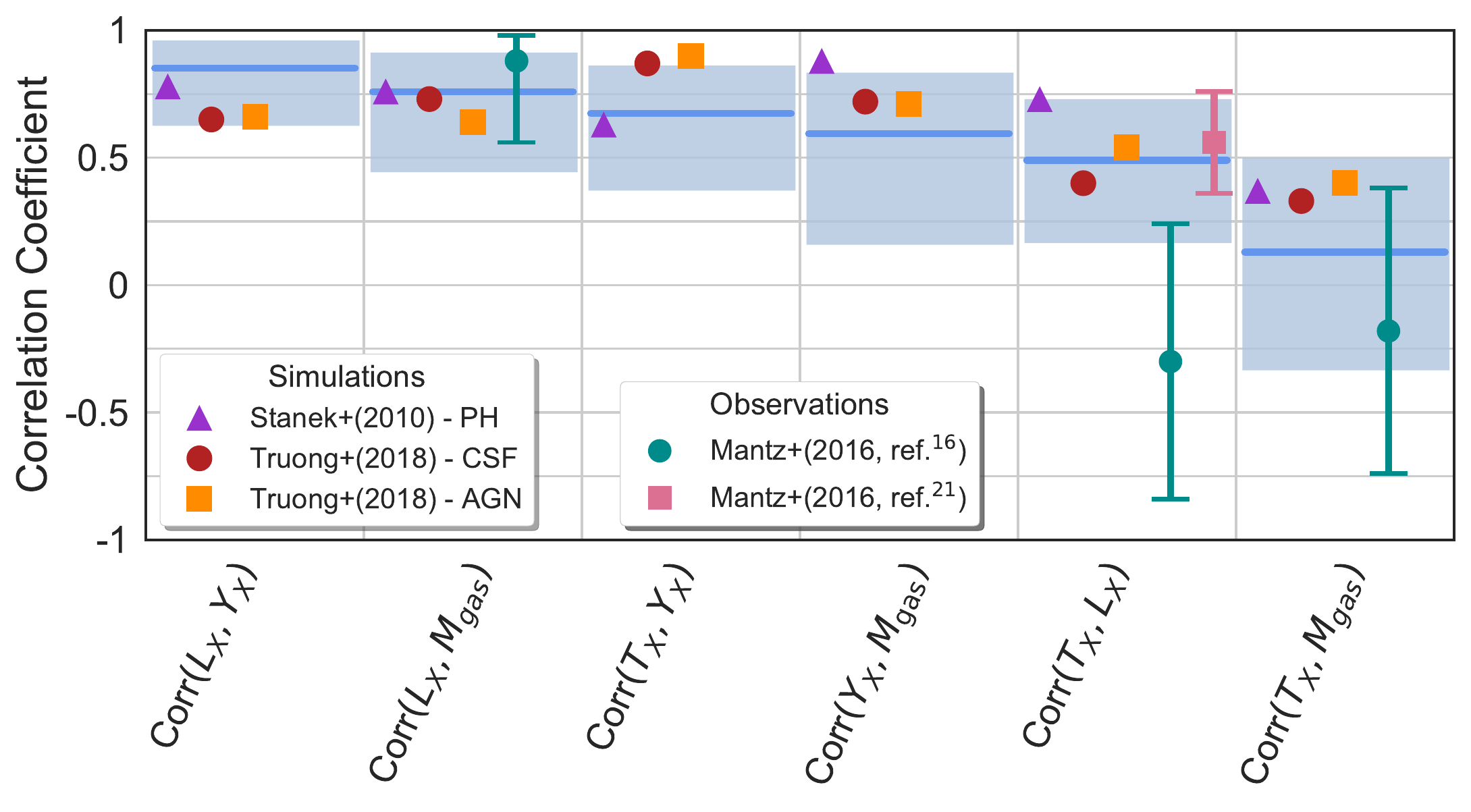} 
 \caption{ Correlation coefficients of X-ray property pairs at fixed halo mass.  Long lines with shaded boxes show the median and posterior 95 percentile constraints from the LoCuSS cluster sample (this work) while filled markers without error bars show expectations published from cosmological hydrodynamical simulations given in the legend. Measurements from previous observational studies given in the legend show central values with 68\% confidence errors.  The axis labels use $L_X$ in place of \Lx \ for simplicity.}
 \label{fig:corr-x-ray}
\end{figure}

Figure~\ref{fig:corr-x-ray} shows posterior constraints of correlations among different hot gas properties.  Our results (shaded bands) of mainly strong positive correlations are consistent with both hydrodynamic simulation expectations \cite{Stanek:2010, Truong:2018} and previous empirical measurements \cite{Mantz:2016Scaling, Mantz:2016b, Andreon:2017}.  Our constraints are broader in scope, i.e. a larger number of measurements of the ICM including $Y_{X}$, $Y_{\rm SZA}$, and $Y_{\rm Pl}$, and in most cases, more precise than the few existing estimates.

\section*{Discussion}

Property covariance has been forecast to significantly improve the power of joint, multi-wavelength survey analysis, especially in the case of anti-correlated properties \cite{Cunha:2009}.  This work helps set the stage for such analysis by providing initial estimates of stellar and hot gas covariance and refined estimates of a larger number of property correlations.  While statistical errors in our correlation estimates are currently large, the coming decades will see an explosion of multi-wavelength cluster data from wide-area surveys such as Large Synoptic Survey Telescope (LSST), Euclid, \emph{e-ROSITA}, and the Stage-4 ground-based cosmic microwave background experiment (CMB-S4). These upcoming samples will allow a better understanding of the physics and feedback effects that regulate the ICM along with improved cosmological constraints from joint sample analysis.

For example, the application of cluster gas fraction as a standard ruler\cite{Mantz:2014} could benefit from the inclusion of stellar mass estimates.  Just as lightcurve shape and color corrections are used to improve the quality of Type Ia supernovae distances, so could stellar mass measurements be used to derive a lower scatter distance proxy than provided by the gas fraction alone.

On the computational side, the sensitivity of cluster property population statistics to modeling treatments will become more apparent as computational advances in multi-phase plasma astrophysics enable refinements of processes at sub-resolution scales.  Synthetic observations of halo populations produced along past lightcones under model-specific conditions, when mapped through survey-specific observational filters, offer a pathway for likelihood testing of increasingly sensitive, multi-wavelength observational surveys.  A new era has arrived for studies of clusters of galaxies as a population, one in which astrophysics-dependent population statistics realized from simulations are tested against corresponding multi-wavelength, empirical data, with outcomes driving improvements in next generation models. Our results reinforce the discovery power of applying population statistical analysis to galaxy cluster samples with complete, uniform multi-wavelength observations that probe hot and cold phase baryons and total mass.

\section*{Methods}

Here we briefly describe the  data vector and regression analysis model employed in this study. Further details are provided in the companion paper \cite{Mulroy:inprep} that discusses mean scaling behaviors and property variance. This paper presents off-diagonal property covariance terms, except for \Lxrass\ correlations which are presented in the companion paper.

\noindent {\bf Cosmology and notation.} We assume a universe with dimensionless energy densities at the current time in total matter (baryons plus dark matter) $\Omega_{\rm m}=0.3$ and vacuum energy $\Omega_\Lambda=0.7$, with Hubble constant $H_0=70\,{\rm km\,s^{-1}\,Mpc^{-1}}$. The Hubble expansion rate is normalized via $E(z)\equiv H(z)/H_{0}=\sqrt{\Omega_{\rm m}(1+z)^{3} + \Omega_\Lambda}$.  For the halo population, we employ a mass scale convention, $M_{500}$, defined as the mass within a sphere, of radius $r_{500}$, within which the mean enclosed density is $500 \rho_{\rm crit}(z)$, where $\rho_{\rm crit}(z) = 3H(z)^2/8\pi G$, is the critical density of the universe\cite{Dodelson:cosmo2003}. Unless stated otherwise, the weak-lensing determined radius, $r_{500,\rm WL}$, defines the aperture within which integrated observable properties are derived. 

\noindent {\bf A multi-wavelength vector of observables.} We study 41 X-ray bright clusters of the LoCuSS sample derived from RASS \cite{BCS,EBCS,REFLEX}.  The sample is selected by  redshift-dependent thresholds of X-ray luminosity, $L_{X,\rm RASS}E^{-1}(z) > 4.4\times 10^{44}\,\rm  erg\,s^{-1}$ for clusters between $0.15<z<0.24$ and $L_{X,\rm RASS}E^{-1}(z) > 7.0 \times 10^{44}\,\rm erg\,s^{-1}$ for  $0.24<z<0.30$.  For each cluster nine additional properties, listed in Table~\ref{tab:obs}, have been measured. Details are provided in the companion paper \cite{Mulroy:inprep}. The sample is complete in most, but not all, properties, as detailed below.  The integrated observables can be grouped into three distinct sets: (i) a weak-lensing mass estimate of total system mass; (ii) quantities associated with the hot intracluster gas, and; (iii) quantities associated with stellar properties.   We briefly describe each set as follows.

(i) Weak-Lensing Mass: Using deep, multi-band optical images from Subaru/Suprime-Cam, a mass estimate for each cluster is derived by fitting the shear signal expected from weak gravitational lensing of a projected Navarro–Frenk–White (NFW, ref. \cite{NFW1997}) mass density profile to the measured tangential shear pattern \cite{Okabe:2016}.  

(ii) Hot Gas Content: Properties of the hot gas content of clusters are mostly observable at X-ray and millimeter wavelengths. We use X-ray measurements of the ICM derived in ref. \cite{Martino2014}, where the selected sample has been observed with either or both of the \emph{Chandra} and \emph{XMM-Newton} X-ray observatories. 

To avoid contamination from the complex cool core region, measurements of bolometric luminosity, $L_{X}$, and gas temperature, $T_{X}$, are performed in an annulus of $[0.15-1]r_{500,\rm WL}$.  The gas mass, $M_{\rm gas}$, is estimated from the observed X-ray emission profile within $r_{500,\rm WL}$. The SZ effect, caused by the inverse Compton scattering of cosmic microwave background (CMB) photons by hot electrons in the ICM\cite{SZ:1980}, is characterized by the parameter $Y$, which is proportional to the integrated electron thermal energy.  The SZ effect from CMB intensity maps, $Y_{\rm SZ}$, is measured via interferometry with SZA and independently with spectral filtering of Planck satellite data. A third estimate of the integrated electron thermal energy, $Y_{X}$, is derived from the X-ray observations as the product of gas mass, $M_{\rm gas}$, and temperature, $T_{X}$.  This quantity is measured within its own iteratively-defined $r_{500}$, as discussed in ref. \cite{Mulroy:inprep}.

(iii) Stellar Content: We employ two independent measures of the stellar content of clusters, the total near-infrared (NIR) luminosity, $L_K$, and a count of red-sequence galaxies, $\lambda$, referred to as optical richness.  The NIR luminosity measurements, obtained with the WFCAM instrument on the UKIRT telescope \cite{Mulroy2014}, determine the background-subtracted light within the weak-lensing estimated radius, $r_{500}$, of each cluster, as well as $L_K$ of the BCG. NIR data is missing for one cluster (Abell2697).  The optical richness, $\lambda$, a measure of the number of red-sequence galaxies within the cluster used by the redMaPPer cluster detection algorithm\cite{Rykoff:2014}, is determined for 33 clusters in the overlap region of the LoCuSS sample and the Sloan Digital Sky Survey (SDSS, ref. \cite{Alam2015}).

\noindent {\bf Regression model.} We assume a log-normal probability distribution of cluster properties with mean values that scale as a power-law in halo mass and $E(z)$.  Because of the narrow redshift range of the LoCuSS sample, we assume standard, self-similar evolution in redshift.  We employ a hierarchical Bayesian inference model that accounts for the sample selection truncation, measurement error covariance and intrinsic property covariance.  An additional component of this inference model is a prior function on true halo masses derived from the halo mass function in the reference $\Lambda$CDM cosmology with $\sigma_{8} = 0.8$.  The performance of this method to recover input scaling relations of synthetic, truncated samples is demonstrated in the companion paper \cite{Mulroy:inprep}.

The key element of our model is the conditional joint property likelihood\cite{Evrard:2014}, $p(\mathbf{S} \,|\, \mhalo,z)$, of a vector of observables, $\mathbf{S}$ (elements in Table~\ref{tab:obs}), given the true mass of the halo, $\mhalo$, at redshift, $z$.  For the LoCuSS sample clusters, we assume that the cluster weak-lensing mass, $M_{WL}$, is an unbiased measure of $\mhalo$ with 20\% fractional scatter.  Our method returns posterior estimates of the intercepts, slopes, and intrinsic variance of each property element as a function of the cluster weak-lensing mass, along with the covariance of pairs of observables. The latter is assumed to be independent of mass and redshift within the narrow ranges probed by the LoCuSS sample. Uninformative priors are used in the analysis.

Using natural logarithms of the properties, $\mathbf{s} = \ln\mathbf{S}$, and mass, $\mu = \ln \mhalo$, the log-mean scaling of observable $a$ at a fixed redshift is linear
 \begin{equation} \label{eq:scalingmodel}
 \langle s_a \, | \, \mu, z \rangle \ = \ \pi_a + \alpha_a \mu \ , 
\end{equation}
in which $\alpha_a$ and $\pi_a$ are the slope and normalization of the scaling relation of property $a$. 

For a pair of observables, $a$ and $b$, the intrinsic property covariance matrix is 
\begin{equation} \label{eq:cov-estimator}
C_{a,b} \ = \ \frac{N}{N-1} \sum\limits_{i=1}^{n} ~ \delta s_{a,i} ~ \delta s_{b,i} \ ,
\end{equation}
where $\delta s_{a,i} \equiv s_{a,i}- \alpha_a \mu_i - \pi_a$ is the residual deviation from the mean scaling relation and $N$ is the total number of clusters. Finally the property correlation coefficient is 
\begin{equation} \label{eq:r-estimator}
r_{a,b} \ = \ \frac{ C_{a,b}}{\sqrt{C_{a,a} \ C_{b,b}} } \ .
\end{equation}
This correlation coefficient is the quantity of interest that is studied in this letter.  Our method constrains these correlation coefficients and the scaling parameters simultaneously, while including a covariance contribution from the reported measurement errors of the properties.

\noindent {\bf Statistical significance and scatter in K-band luminosity.}
In the companion paper, we show that the posterior constraints on the intrinsic scatter in \Lk\  are not bounded from below; values near zero are not only allowed by the data but the modal value of the posterior Probability Density Function (PDF) is zero.  The correlation coefficients between \Lk\ and other properties vary substantially as the scatter in $\ln \,$\Lk \ drops to very low values.  Very small values of this scatter, $\sigma_{\ln L_K|M}$, are not physically reasonable.   
Cosmological hydrodynamics simulations have found values of $\sigma_{\ln L_K|M}>0.10$\cite{Farahi:2017bahamas}, $0.32$\cite{Wu:2015}, or $0.16$\cite{Pillepich:2018}; and a recent observational study estimates a value of $0.22 \pm 0.04$\cite{Chiu:2018}.

The confidence intervals and statistical significance of the anti-correlation signals reported in the main text employ a lower limit of $\sigma_{\ln L_K| M} = 0.05$.  This choice is a  conservative one, two times smaller than the smallest value reported above. We discard any point in the posterior chain with $\sigma_{\ln L_K| M} < 0.05$. All numbers reported in the main text are based on this truncated posterior distribution. For the sake of symmetry, we also impose the same limit on the richness scatter, $\sigma_{\ln \lambda| M} \ge  0.05$, but this has a much smaller effect as the posterior PDF has very little support in this region.

\begin{figure}
  \centering
  \includegraphics[width=0.6\linewidth]{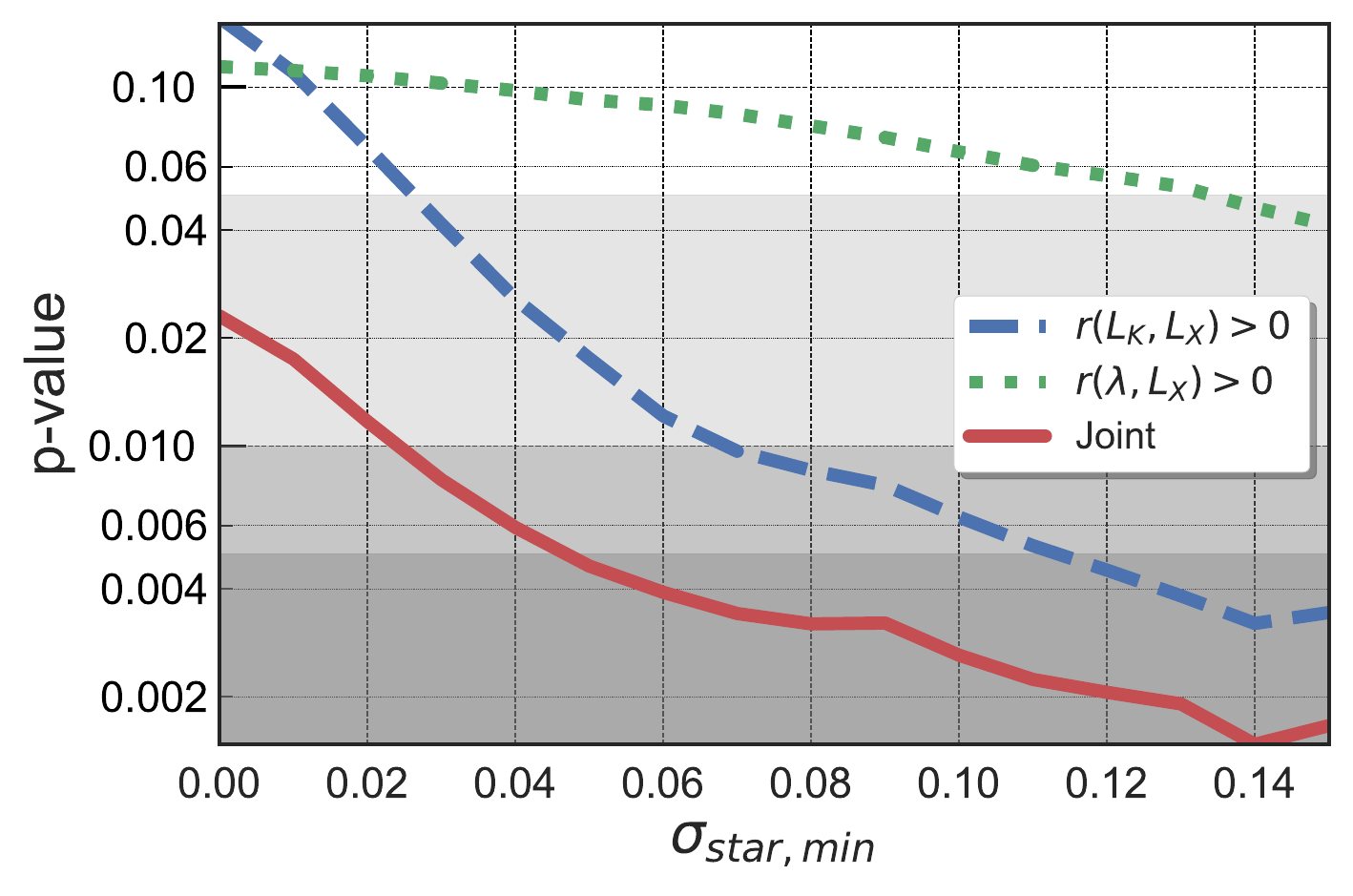} 
 \caption{ Odds that the stellar--$L_X$ correlation is positive as a function of the minimum imposed stellar mass scatter. The blue dashed and green dotted lines shows the odds for stellar mass proxies of $L_K$ and $\lambda$, respectively. The red solid line shows the odds of having a positive correlation coefficient for both stellar mass proxies and $L_X$. The shaded regions shows the odds of 5\%, 1\%, and 0.5\%.}
 \label{fig:p-val-test}
\end{figure}

The statistical significance of the hot-cold baryon phase anti-correlation reported here is sensitive to the choice of minimum value for the stellar mass scatter.  Figure~\ref{fig:p-val-test} illustrates the odds of having a positive correlation for an optical and $L_X$ observable changes as a function of the imposed minimum value of $\sigma_{\ln M_{\rm star}| M}$.  For a minimum value of 0.1, the odds of a positive correlation between $L_K$ and $L_X$ at fixed halo mass are 0.006, or roughly three-$\sigma$ evidence.  The odds that {\sl both} optical measures correlate positively with \Lx\ is very small, $0.005$ for our fiducial minimum of $0.05$ in $\sigma_{\ln M_{\rm star}| M}$.

\section*{Data availability}

The observational data vector employed in this work is available via the companion paper\cite{Mulroy:inprep} [\url{https://academic.oup.com/mnras/article/484/1/60/5274143}]. The full posterior chains that support the findings of this study is available in a figshare repository,  [\url{https://doi.org/10.6084/m9.figshare.8001218}]. The source data underlying Figs 1, 2, and 3 and Table 2 are reproducible via the posterior chains.

\section*{Code Availability}

All analyses were performed with custom-built Python scripts. The MCMC is performed with \texttt{PyMC}\cite{pymc} [\url{https://pymc-devs.github.io/pymc/}], a Bayesian stochastic modeling package in Python. The analyses scripts are available from the corresponding author upon request.

\bibliography{sample}

\section*{Acknowledgements}

We are grateful to Adam Mantz and Ben Maughan for useful discussions on population statistics. We thank Arif Babul and members of the LoCuSS, DES and XMM-XXL collaborations for their support. A.F. is supported by a McWilliams Postdoctoral Fellowship. A.E.E. and A.F. acknowledge support from NASA Chandra grants GO8-19107B and GO6-17116B. S.L.M. and G.P.S. acknowledge support from the STFC. G.P.S. acknowledges support from the Royal Society. C.O., D.P.M., Z.A., J.E.C., and CARMA operations were supported by NSF grant AST-1140019. C.P.H. acknowledges support from PRIN INAF 2014. H.B. and P.M. acknowledge financial contribution from the agreement ASI-INAF n.2017-14-H.O and from University of Rome ``Tor Vergata'' Grant, ``Mission: Sustainability'' EnClOS.

\section*{Author contributions statement}

 A.F. led the data analysis with contributions from A.E.E., S.L.M. G.P.S., and A.F. S.L.M. and G.P.S led the multi-wavelength observational studies. A.F. implemented, tested, and performed the statistical analysis. A.F. and A.E.E wrote the main body of the manuscript. H.B., R.M. and P.M. contributed in deriving all the observational quantity related to the X-ray observations of the hot gas content and the SZ signal with the Planck satellite. C.P.H. was responsible for the K-band photometry and cluster galaxy selection. N.O. was responsible for measuring the weak-lensing masses.
 D.P.M. and J.E.C. were responsible for obtaining the SZA observations and D.P.M supervised their analysis. C.O. analyzed interferometric data from SZA to estimate the $Y_{SZ}$ for each cluster. All authors discussed the results, contributed to the discussions in the manuscript, and reviewed the manuscript. G.P.S. proposed and designed the LoCuSS (the principle investigator).

\section*{Competing interests}

The authors declare no competing interests.

\section*{Supplementary Information}

\subsection*{Systematic Effects}

 In this section, we study the systematic effects to support the claims made in the main text. We are primarily interested in constraining the correlation coefficient between hot gas mass and stellar mass of the underlaying halos population, which is expected to be anti-correlated. We examine here sources of bias which could modify the estimated correlation. Throughout this section we assume power-law for the mean relation between mass and observables with log-Normal scatter, unless otherwise mentioned.

 \begin{figure}
  \centering
  \includegraphics[width=0.7\linewidth]{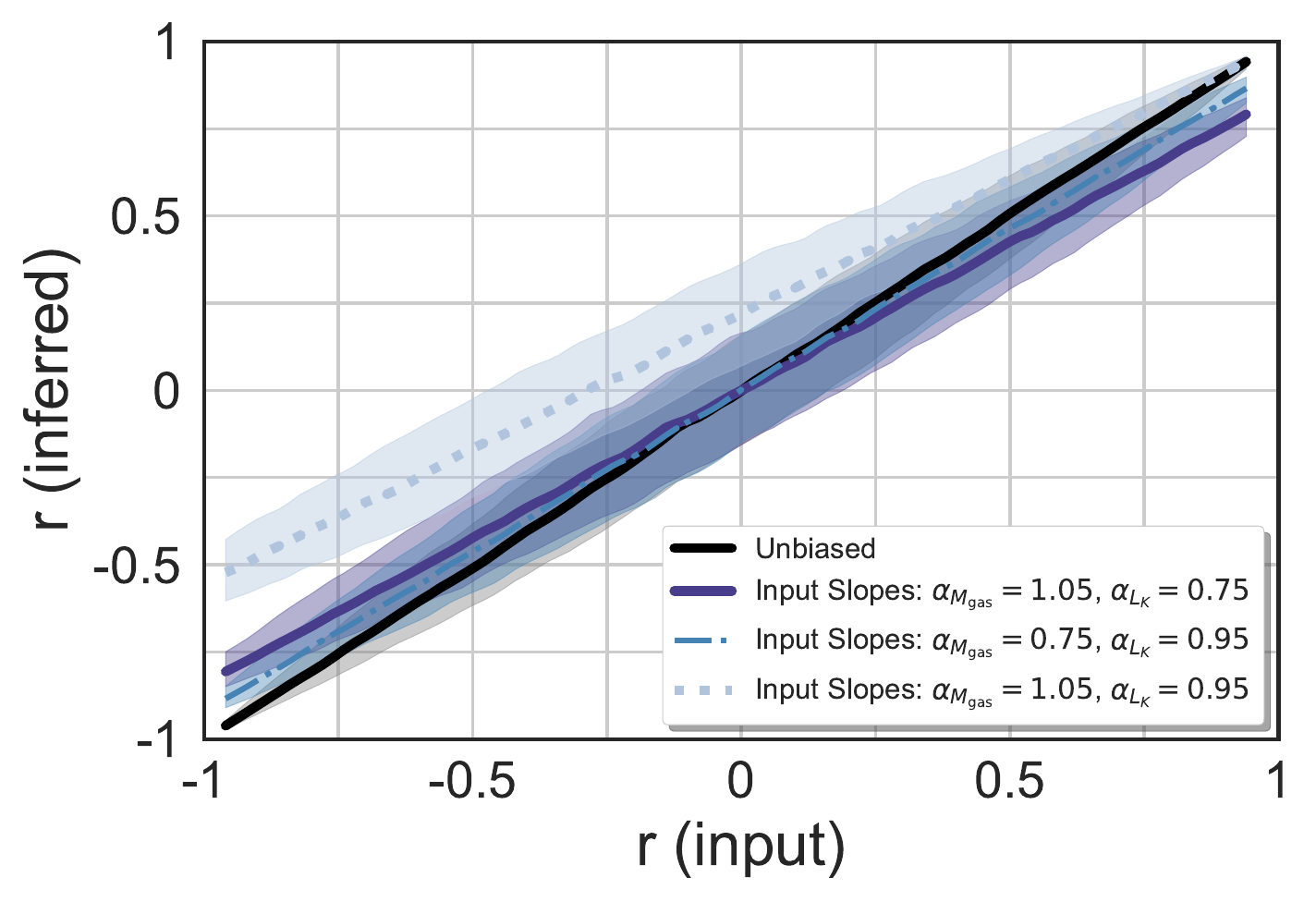} 
 \caption{  The effect of underestimating the slope of the scaling relation on the estimated property correlation.  Sensitivity test of the estimated property correlation for synthetic samples patterned after the LoCuSS cluster sample. A posterior slope of 0.75 for both \Mgas--\Mwl\ and \Lk--\Mwl\ relations is assumed, while the input true slope takes the values specified in the legend. The shaded areas are $68\%$ confidence intervals derived from 1,000 realizations for each input correlation coefficient. }
 \label{fig:corr-test-1}
\end{figure}

\subsection*{Effect of biased scaling relation slope} 

One potential systematic error arises from a potential bias in the posterior mean scaling relation parameters relative to the underlying truth. To illustrate this effect, we generate a synthetic sample of clusters with a known input scaling parameters, then estimate the correlation coefficient imposing a biased set of parameters. To start, we take the LoCuSS weak-lensing masses and assume a set of \Mgas--\Mwl\ and \Lk--\Mwl\ relations. For each cluster, we draw a random \Mgas\ and \Lk\ from a multivariate log-normal distribution with a set of input correlation coefficients and $20\%$ intrinsic scatter. Cluster residuals about the mean relation are estimated assuming a set of biased scaling relations consistent with ref.\cite{Mulroy:inprep}. Finally, the correlation coefficient is estimated with Equation \ref{eq:r-estimator}. For each input correlation coefficient, 1,000 realizations of LoCuSS-like cluster sample are generated. Figure~\ref{fig:corr-test-1} illustrates the shift in the estimated correlation coefficient as a function of input correlation coefficient. 
 
 While the correlation coefficients based on the gas mass, \Mgas, support negative values for both \Lk\ and \richness, but the statistical significance is smaller than X-ray luminosity, \Lx, and stellar mass proxies. In the companion paper, we note that the slope of the \Mgas\ scaling with halo mass is $0.77\pm0.1$, lower than values above one derived by previous observational studies\cite{Mantz:2016Scaling} and from modern hydrodynamic simulations\cite{Farahi:2017bahamas}, which yield values slightly above unity. The purple line in Figure \ref{fig:corr-test-1} shows that a bias of 0.3 in $\mgas$ slope would produce an underestimate in the anti-correlation magnitude, potentially helping to explain why \Lx\ provides more significant evidence for anti-correlation than \Mgas.

\subsection*{Effect of extrinsic or underestimated statistical errors} 
Our likelihood model assumes that statistical errors of each property are accurate and that the remaining residuals about the mean scaling relation reflect only intrinsic scatter of the underlying halo population that host the clusters in the LoCuSS sample. Properties that are subject to extrinsic contributions not already incorporated into the statistical error budget represent another potential source of systematic error. For example, a recent study shows that the uncertainties quoted for \richness\ by the redMaPPer algorithm underestimate the extrinsic scatter driven by projected galaxies lying outside the host halo\cite{Costanzi:2018}.

\begin{figure}
  \centering
  \includegraphics[width=0.7\linewidth]{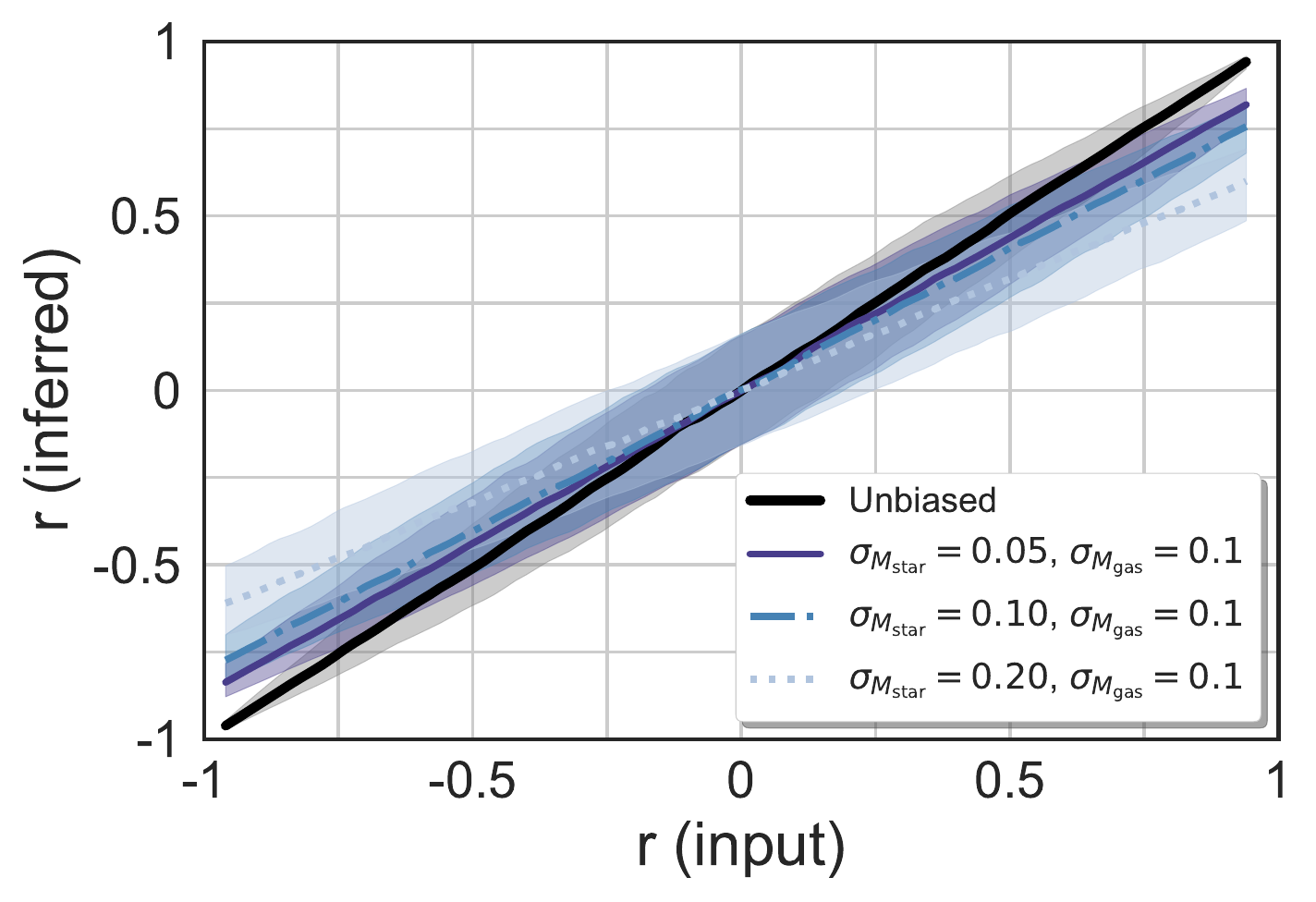} 
 \caption{ The effect on the estimated property correlation of extrinsic scatter. Similar to Figure~\ref{fig:corr-test-1}, we generate synthetic realizations 41 pairs of intrinsic quantities -- $\mgas$ and $\mstar$ -- with a known property covariance. We then perturbed intrinsic quantities with extrinsic scatter to obtain measured quantities. The magnitude of extrinsic scatter is specified in the legend. Shaded areas are $68\%$ confidence intervals derived from 1,000 realizations for each input correlation coefficient.   }
 \label{fig:corr-test-2}
\end{figure}

We test the effect of such systematics by realizing synthetic samples of pairs of properties, \{\Mstar, \Mgas\}, subject to additional, uncorrelated sources of uncertainty.  We assume \Mstar--\Mwl\ and \Mgas--\Mwl\ relations with unit mean slopes and a two-dimensional, log-normal probability distribution with $20\%$ intrinsic scatter in each component and an input correlation coefficient, $r$.  Each halo property is then further perturbed with uncorrelated scatter in each component, of variable magnitude $\sigma_{\rm star}$ and $\sigma_{\rm gas}$, to obtain the observed quantities. For each input correlation, we generate 1,000 realizations of 41-object samples, measuring the property correlation between the observed quantities with our likelihood model. Figure \ref{fig:corr-test-2} illustrates that such extrinsic scatter leads to inferred correlations that underestimate the magnitude of the underlying population. We note that the difference in posterior correlation coefficient estimates for the $\{\mgas,$ \Lk $\}$ pair and $\{ \mgas , \lambda \}$ pair seen in Figure~\ref{fig:corr-x-ray-optical} could be understood if \richness\ is subject to a larger extrinsic scatter effect than \Lk.

The effects of bias in inferred slope and uncalibrated scatter can explain the systematic shifts in the marginalized posterior distributions seen in Figure \ref{fig:corr-x-ray-optical}.  With current data, we are not able to assess the contribution of each systematic effect. For example, whether the richness is a noisier measurement of true halo stellar mass with respect to \Lk, or whether the richness measurement uncertainties are underestimated, or both, cannot be addressed with the existing observational data. Detailed simulation and future observational studies are needed to calibrate and understand these effects. 

\subsection*{Orientation effects} 

Massive halos deviate from spherical symmetry, with collisionless components tending toward prolate ellipsoidal shapes with minor to major axis ratio of 0.6 \cite{Kasun:2005}.  The collisional nature of the hot gas drives it to a more spherical shape formed by equipotential surfaces. It is reasonable to ask whether correlations between weak-lensing mass and stellar properties driven by orientation might induce an anticorrelation with hot gas properties.  To address this question, we apply the log-normal population model of \cite{Evrard:2014} under the assumption of moderate to strong covariance between weak-lensing mass and total K-band light at fixed true mass, and zero intrinsic correlation of either of these properties with hot gas properties.  We show here that the correlation of hot gas and K-band luminosity conditioned on weak-lensing mass can be driven to negative values, but only under extreme values of model parameters that are very unlikely.

We consider the likelihood of a vector of properties, $p(\mathbf{s} \, | \, M_{\rm true},z)$, with elements, $\mathbf{s} = \ln  \{M_{\rm WL}, L_k, X \}$ ($X = $ a hot gas property such as $L_X, M_{\rm gas}$, etc.)  and let $\mu = \ln M_{\rm true}$. The PDF of true selected mass, $p(\mu \, | \, M_{\rm WL})$, is Gaussian with an assumed width $0.2$. Considering the optical properties, $L_K$ or $\lambda$, for a sample selected on weak-lensing mass, the correlation between $\lambda$ (for example) and true halo mass, $\mu$, at fixed weak-lensing mass is given by equation (8) of ref.\cite{Evrard:2014}, 
\begin{equation} \label{eq:r-analytic}
r_{\ln \lambda ,\mu \, | \, M_{\rm WL}}  = \frac{ \sigma_{\mu \,|\, M_{\rm WL}} / \sigma_{\mu \,|\, \lambda} - r} { \left( 1-r^2 + (\sigma_{\mu \,|\, M_{\rm WL}} / \sigma_{\mu \,|\, \lambda} -r)^2 \right)^{1/2} },
\end{equation}
where $\sigma_{\mu \,|\, M_{\rm WL}} = 0.2$ is the scatter in true mass at fixed weak-lensing mass, $\sigma_{\mu \,|\, \lambda}$ is the similar measure at fixed $\lambda$, and $r$ is the correlation coefficient of $\ln M_{\rm WL}$ and $\ln \lambda$ at fixed $M_{\rm true}$. Similar expressions apply for $\lambda \rightarrow L_K$.  We assume that the correlation coefficient of weak-lensing-mass and richness is significant, $r \gtrsim 0.5$. The above expression indicates that an anti-correlation between true mass and $\lambda$ or $L_K$ could be induced if the numerator is negative.  With the default assumption that a hot gas property, $X$, is uncorrelated with $M_{\rm WL}$ and $\lambda$ at fixed true mass, $\mu$, this condition would produce an anti-correlation of $\lambda$ and $X$ at fixed $M_{\rm WL}$, consistent with our findings.  

However, the requirement that Equation \ref{eq:r-analytic} be negative implies that the mass scatter at fixed K-band total luminosity must be large in order to suppress the first term in the numerator, 
\begin{equation}
\sigma_{\mu \,|\, \lambda}  \ge  \sigma_{\mu \,|\, M_{\rm WL}} / r  = 0.2 / r .  
\end{equation}
A previous work that studied orientation and projection effects of different mass proxies in N-body simulations finds $r = 0.55$ for weak-lensing and red galaxy count\cite{NohCohn:2012}, meaning $\sigma_{\mu \,|\, \lambda} \simeq 0.4 $ .

For this effect to be responsible for our findings of a large anti-correlation coefficient between hot and cold gas properties, two factors must conspire: i) $M_{\rm WL}$ and $L_K$ (or $\lambda$) at fixed true mass must be very tightly tied, $r \sim 0.9$, and, ii) both optical proxies must have a factor $\sim 2$ larger scatter in selected true mass compared to $M_{\rm WL}$. Such a strong correlation between weak-lensing mass and optical richness is unlikely, given that random LSS projections across a broad redshift range contribute at least $0.06$ to the scatter in $M_{\rm WL}$ \cite{Hoekstra:2001}.  Such a large intrinsic scatter in optical richness for halos above $5 \times 10^{14} \msol$, above which the majority of LoCuSS clusters lie, is not supported by state-of-the-art hydrodynamics simulations\cite{Farahi:2017bahamas} and existing observational studies\cite{RozoRykoff:2014}. 

The following section includes an explicit test that confirms the required conditions.

\subsection*{Line-of-sight projection effect}

The cluster properties we use are integrated quantities in the projected space of sky coordinates and photon frequency.  Projection of extrinsic material will perturb the intrinsic quantities and, unlike the extrinsic study above, can do so in a correlated manner\cite{NohCohn:2012}. Here we model the effect of projection explicitly as a source of extrinsic covariance, $\Sigma_{\rm total} = \Sigma_{\rm halo} + \Sigma_{\rm proj}$. The covariance estimated in this work, Table \ref{tab:corr-post}, is the total covariance, and our findings in the main text implicitly assume $\Sigma_{\rm total} \approx \Sigma_{\rm halo}$. Because $\Sigma_{\rm halo}$ and $\Sigma_{\rm proj}$ are degenerate, we cannot disentangle these two components with the observational data in hand.

We perform a set of explicit simulations similar to what we have done above. We generate synthetic realizations of intrinsic quantities of a set of halos with an input property covariance, specified in Table \ref{tab:proj-covariance} (a), and four models for projection covariance  specified in Table \ref{tab:proj-covariance} (b)-(e). Model (b) is motivated by the fact that the weak-lensing mass and optical observables are expected to be highly correlated \cite{NohCohn:2012}. Model (c) is similar to model (b) with larger extrinsic scatter, which illustrates the diluting effect of such a large scatter. It is expected that the optical and weak-lensing observables to be correlated with hot gas observables, but smaller in magnitude \cite{NohCohn:2012}. Thus, we consider a third model -- model (d) -- which includes a small correlation between the X-ray observable and the optical and the mass observables.  Finally, model (e) is motivated by our analytical model of the previous section. This improbable model shows that a large extrinsic optical scatter strongly coupled with the mass proxy may indeed induce negative covariance.

Shaded areas are $68\%$ confidence intervals for inferred correlation derived from 1,000 realizations for each input correlation coefficient.  The set examines a broad range of scenarios for the projection effect, and our results confirm that negative correlations are diluted. Unless the scatter due to the 2-halo term (projection effect) and correlation between $M_{\rm WL}$ and $L_K$ is very large the projection cannot induce a negative correlation. 

As a final test, we revise our inference algorithm to include an explicit projection term with redefining the covariance $\Sigma = \Sigma_{\rm int} + \Sigma_{\rm proj}$. Due to the degeneracy between $\Sigma_{\rm int}$ and $\Sigma_{\rm proj}$, we cannot constrain both quantities simultaneously. We, therefore, employ a fixed $\Sigma_{\rm proj}$, model parameters specified in Table \ref{tab:proj-covariance} (b), and preform the likelihood to infer $\Sigma_{\rm int}$. We find that $\Sigma_{\rm int}$ posteriors are consistent with our main results. Saying that, it is worth noting that the statistical significance of the anti-correlation between gas mass and K-band luminosity is improved -- new p-value $= 0.03$, and new $r=-0.70 ^{+0.34}_{-0.20}$.

 \begin{figure}
  \centering
  \includegraphics[width=0.7\linewidth]{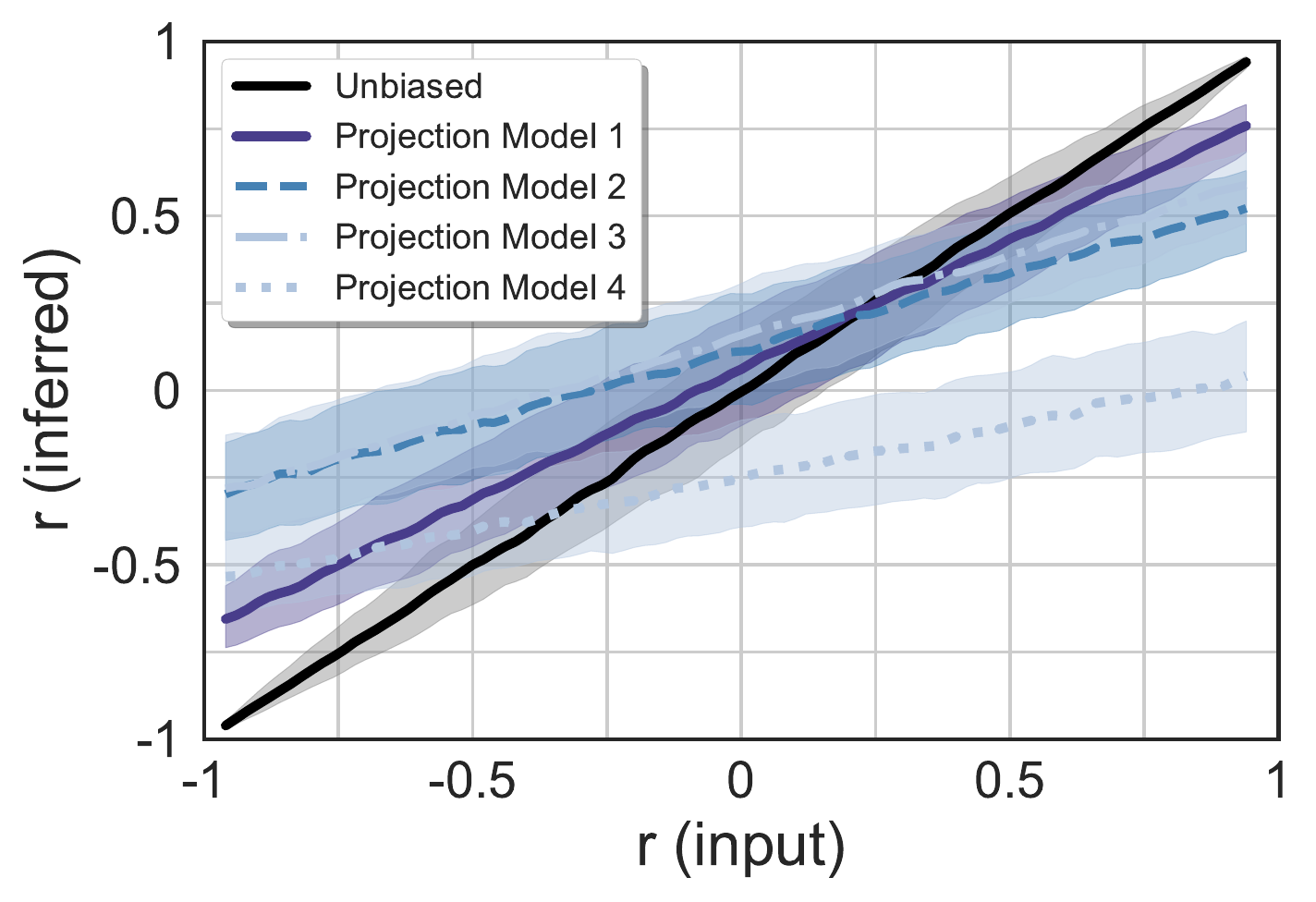} 
 \caption{  The effect of projection effect covariance on the estimated property correlation.  We generate synthetic realizations of intrinsic quantities of a set of halos with an input property covariance -- specified in Table \ref{tab:proj-covariance} (a) -- and a projection covariance -- specified in Table \ref{tab:proj-covariance} (b)-(e) --  to obtain measured quantities. The additional random scatter is specified in the legend.  Shaded areas are $68\%$ confidence intervals derived from 1,000 realizations for each input correlation coefficient.  }
 \label{fig:projection-test}
\end{figure}

We do not attempt to correct for this effect as an estimation of such effect is poorly constrained. Estimating the magnitude of this effect requires comprehensive numerical simulations, making synthetic observations, and performing the measurement algorithms which is beyond the scope of this work and is a subject of our future studies. We note that for a larger sample with better statistical power, this effect need to be calibrated and better understood.

\subsection*{Systematics Test Conclusion}

The main conclusion drawn from the above tests is that these systematics cannot not induce an anti-correlation signal. Instead, any systematics which is not accounted for or exists in our dataset would reduce the statistical significance of our results and dilute the intrinsic halo anti-correlation signal. Therefore, our results should be interpreted as a conservative lower bound on the magnitude of the anti-correlation between $\mgas$ and $\mstar$ at fixed halo mass. As the quality of data improving and the sample size growing, we need to handle the above systematics better. A future direction should, particularly, address the projection effect for a broad range of observables.

\begin{table}[!htb]
    \caption{The projection covariance models employed in the Supplementary Material to assess the effect of the projections on the inferred anti-correlation signal. The diagonal elements are the scatter and the off-diagonal elements are the correlation coefficients.} \label{tab:proj-covariance}
    \begin{subtable}{\linewidth}
      \centering
      \caption{Input -- $\Sigma_{\rm input}$}
		\begin{tabular}{|l|c|c|c|}
			\hline
			\rowcolor[HTML]{A9A9A9} \cellcolor[HTML]{333333} & Mass & Optical                  & X-ray                                           \\ \hline
			\rowcolor[HTML]{A9A9A9} Mass                     & \cellcolor[HTML]{CCCCCD} 0 & \cellcolor[HTML]{333333} & \cellcolor[HTML]{333333}                        \\ \hline
			\rowcolor[HTML]{A9A9A9} Optical                  &  \cellcolor[rgb]{1.000, 0.622, 0.460} 0  & \cellcolor[HTML]{CCCCCD} 0.15                     & \cellcolor[HTML]{333333}{\color[HTML]{000000} } \\ \hline
			\rowcolor[HTML]{A9A9A9} X-ray                    &  \cellcolor[rgb]{1.000, 0.622, 0.460} 0    &  \cellcolor[rgb]{1.000, 0.622, 0.460} r (input)                        & \cellcolor[HTML]{CCCCCD} 0.15                                            \\ \hline
		\end{tabular}
    \end{subtable}   \\
    \\
    
    \begin{subtable}{.48\linewidth}
      \centering
      \caption{Projection Model 1  -- $\Sigma_{\rm projection, 1}$}
		\begin{tabular}{|l|c|c|c|}
			\hline
			\rowcolor[HTML]{A9A9A9} \cellcolor[HTML]{333333} & Mass & Optical                  & X-ray                                           \\ \hline
			\rowcolor[HTML]{A9A9A9} Mass                     & \cellcolor[HTML]{CCCCCD} 0.10 & \cellcolor[HTML]{333333} & \cellcolor[HTML]{333333}                        \\ \hline
			\rowcolor[HTML]{A9A9A9} Optical                  &  \cellcolor[rgb]{1.000, 0.622, 0.460} 0.50 & \cellcolor[HTML]{CCCCCD} 0.10                     & \cellcolor[HTML]{333333}{\color[HTML]{000000} } \\ \hline
			\rowcolor[HTML]{A9A9A9} X-ray                    &  \cellcolor[rgb]{1.000, 0.622, 0.460} 0.00 &  \cellcolor[rgb]{1.000, 0.622, 0.460} 0.00                     & \cellcolor[HTML]{CCCCCD} 0.05                                            \\ \hline
		\end{tabular}
    \end{subtable}%
    \begin{subtable}{.48\linewidth}
      \centering
      \caption{Projection Model 2  -- $\Sigma_{\rm projection, 2}$}
		\begin{tabular}{|l|c|c|c|}
			\hline
			\rowcolor[HTML]{A9A9A9} \cellcolor[HTML]{333333} & Mass & Optical                  & X-ray                                           \\ \hline
			\rowcolor[HTML]{A9A9A9} Mass                     & \cellcolor[HTML]{CCCCCD} 0.20 & \cellcolor[HTML]{333333} & \cellcolor[HTML]{333333}                        \\ \hline
			\rowcolor[HTML]{A9A9A9} Optical                  & \cellcolor[rgb]{1.000, 0.622, 0.460} 0.50 & \cellcolor[HTML]{CCCCCD} 0.20                     & \cellcolor[HTML]{333333}{\color[HTML]{000000} } \\ \hline
			\rowcolor[HTML]{A9A9A9} X-ray                    & \cellcolor[rgb]{1.000, 0.622, 0.460} 0.00 & \cellcolor[rgb]{1.000, 0.622, 0.460} 0.00                     & \cellcolor[HTML]{CCCCCD} 0.10                                            \\ \hline
		\end{tabular}
    \end{subtable}  \\
    \\
    
    \begin{subtable}{.48\linewidth}
      \centering
      \caption{Projection Model 3  -- $\Sigma_{\rm projection, 3}$}
		\begin{tabular}{|l|c|c|c|}
			\hline
			\rowcolor[HTML]{A9A9A9} \cellcolor[HTML]{333333} & Mass & Optical                  & X-ray                                           \\ \hline
			\rowcolor[HTML]{A9A9A9} Mass                     & \cellcolor[HTML]{CCCCCD} 0.20 & \cellcolor[HTML]{333333} & \cellcolor[HTML]{333333}                        \\ \hline
			\rowcolor[HTML]{A9A9A9} Optical                  & \cellcolor[rgb]{1.000, 0.622, 0.460} 0.50 & \cellcolor[HTML]{CCCCCD} 0.20                     & \cellcolor[HTML]{333333}{\color[HTML]{000000} } \\ \hline
			\rowcolor[HTML]{A9A9A9} X-ray                    & \cellcolor[rgb]{1.000, 0.622, 0.460} 0.25 & \cellcolor[rgb]{1.000, 0.622, 0.460} 0.25                     & \cellcolor[HTML]{CCCCCD} 0.10                                            \\ \hline
		\end{tabular}
    \end{subtable}  %
    \begin{subtable}{.48\linewidth}
      \centering
      \caption{Projection Model 4  -- $\Sigma_{\rm projection, 4}$}
		\begin{tabular}{|l|c|c|c|}
			\hline
			\rowcolor[HTML]{A9A9A9} \cellcolor[HTML]{333333} & Mass & Optical                  & X-ray                                           \\ \hline
			\rowcolor[HTML]{A9A9A9} Mass                     & \cellcolor[HTML]{CCCCCD} 0.20 & \cellcolor[HTML]{333333} & \cellcolor[HTML]{333333}                        \\ \hline
			\rowcolor[HTML]{A9A9A9} Optical                  & \cellcolor[rgb]{1.000, 0.622, 0.460} 0.75 & \cellcolor[HTML]{CCCCCD} 0.40                     & \cellcolor[HTML]{333333}{\color[HTML]{000000} } \\ \hline
			\rowcolor[HTML]{A9A9A9} X-ray                    & \cellcolor[rgb]{1.000, 0.622, 0.460} 0.00 & \cellcolor[rgb]{1.000, 0.622, 0.460} 0.00                     & \cellcolor[HTML]{CCCCCD} 0.10                                            \\ \hline
		\end{tabular}
    \end{subtable}  
\end{table}

\end{document}